\RequirePackage{fix-cm}
\documentclass[smallextended]{svjour3}       % onecolumn (second format)
\smartqed  % flush right qed marks, e.g. at end of proof
\usepackage{graphicx}
\usepackage{amsmath}
\usepackage{float}
\begin{document}

\title{Superconductivity in model cuprate as an S\,=\,1 pseudomagnon condensation%\thanks{Grants or other notes
}


\author{E.V. Vasinovich         \and
        A.S. Moskvin            \and
        Yu.D. Panov             
}


\institute{E.V. Vasinovich \at
              Ural Federal University, Ekaterinburg, Russia 620002 \\
              \email{e.vasinovich@gmail.com}           %  \\
           \and
           A.S. Moskvin \at
              Ural Federal University, Ekaterinburg, Russia 620002
           \and
            Yu.D. Panov \at
              Ural Federal University, Ekaterinburg, Russia 620002
}

\date{Received: date / Accepted: date}

\maketitle

\begin{abstract}
We make use of the S\,=\,1 pseudospin formalism to describe the charge degree of freedom in a model high-$T_c$ cuprate with the on-site Hilbert space reduced to the three effective valence centers, nominally Cu$^{1+,\,2+,\,3+}$. Starting with a parent cuprate as an analogue of the quantum paramagnet ground state and using the Schwinger boson technique we found the pseudospin spectrum and conditions for the pseudomagnon condensation with phase transition to a superconducting state.

\end{abstract}

\section{Introduction}
\label{intro}

Interest to the S\,=\,1 spin models\,\cite{Hamer,Sizanov,Guimaraes,Sousa,Moura} is associated with both the description of strongly anisotropic magnets based on Ni$^{2+}$ (conventional spin S\,=\,1), in particular [Ni(HF$_2$)(3-Clpy)$_4$]BF$_4$ \cite{Manson2012,Wierschem2014} and NiCl$_2\cdot$4SC(NH$_2$)$_2$ \cite{NiCL}, and the description of ``semi-hard-core''\, bosons\,\cite{Diehl} or mixed-valence systems such as the system of charge ``triplets''\, Cu$^{1+,\,2+,\,3+}$ in cuprates or Bi$^{3+,4+,5+}$ in bismuthates\,\cite{Moskvin,Moskvin2011,Moskvin2013,Moskvin2015,Vasinovich2018}.

At variance with s\,=\,1/2 quantum magnets the S\,=\,1 spin systems  are characterized by a more complicated Hamiltonian with emergence of a single-ion anisotropy and biquadratic inter-site couplings that give rise to novel phase states, in particular, the quantum paramagnet and spin-nematic order. Theoretical methods that have proved successful in the study of quantum s\,=\,1/2 magnets such as the spin wave theory\,\cite{Diep}, exact diagonalization\,\cite{Dagotto,Zhitomirsky,Richter}, series expansion\,\cite{Singh} large-N expansion\,\cite{Read}, functional renormalization group\,\cite{Reuther}, Green's function method\,\cite{Siurakshina}, and projected entangled pair states\,\cite{Murg}, for the S\,=\,1 spin systems face various difficulties and they need to be reconsidered. Sufficiently effective in this case is the Schwinger boson method\,\cite{Auerbach1988}, which was realized for different systems with spin S\,=\,1\,\cite{Wang2005,Zhang2013,Pires2015,Pires2016,Pires2018}. In our work, this method is generalized for more complicated pseudospin systems which take into account the effects of the biquadratic inter-site anisotropy.

Below we dwell on an analysis of the elementary excitations of the S\,=\,1 pseudospin system or pseudomagnons on a square lattice under the assumption of the quantum paramagnetic ground state. Main interest in the work is initiated by the problem of inducing high-temperature superconductivity in parent cuprates, whose ground state corresponds to valency of copper Cu$^{2+}$, that within the framework of the pseudospin formalism corresponds to the state of the quantum paramagnet   $\langle S_z\rangle$\,=\,$\langle S_z^2\rangle$\,=0.

\section{Charge triplet model}
\label{sec:1}

Hereafter we consider a model 2D pseudospin system of the ``semi-hard-core'' bosons type with a constraint of the lattice site occupancy n = 0, 1, 2, or mixed-valence ion systems of the charge ``triplet'' Cu$^{1+,2+,3+}$ type in cuprates or Bi$^{3+,4+,5+}$ in bismuthates\,\cite{Moskvin,Moskvin2011,Moskvin2013,Moskvin2015}. The simplified charge triplet model implies a full neglect of spin and orbital degrees of freedom. Three charge states of the CuO$_4$ cluster in CuO$_2$ planes (nominaly  Cu$^{2+}$, Cu$^{3+}$, Cu$^{1+}$) are assigned to three components of the S\,=\,1 pseudospin with the projections $M_S = 0, +1, -1$, respectively.
This assignment allows us  to apply well known methods of conventional spin algebra to describe charge degree of freedom in cuprates.

The S\,=\,1 spin algebra includes the eight independent
nontrivial (pseudo)spin operators, the three dipole and five quadrupole ones:
\begin{equation}
S_z;\ S_{\pm} = \mp \frac{1}{\sqrt{2}} (S_x \pm i S_y);\ S_{z}^2;\ T_{\pm} = \{ S_z, S_\pm \};\ S_{\pm}^2.
\end{equation}
The raising/lowering operators $S_{\pm}$ and $T_{\pm}$  change the pseudospin projection by $\pm 1$ with slightly different properties: $ \langle 0 | S_{\pm} | \mp 1 \rangle = \langle \pm 1 | S_{\pm} | 0 \rangle = \mp 1$, $ \langle 0 | T_{\pm} | \mp 1 \rangle = - \langle \pm 1 | T_{\pm} | 0 \rangle = + 1$. These are effective one-particle transfer operators. In lieu of the $S_{\pm}$ and $T_{\pm}$ operators one may use the two novel operators $P_\pm = \frac{1}{2} (S_\pm + T_\pm)$ and $N_\pm = \frac{1}{2} (S_\pm - T_\pm)$, which do realize transformations $|0 \rangle \leftrightarrow |+1\rangle$ and $|0\rangle \leftrightarrow |-1\rangle$. The raising/lowering operators $S_{\pm}^2$ change the (pseudo)spin projection by $\pm$\,2 and describe the $|-1\rangle \leftrightarrow |+1\rangle$ transitions, or the two-particle  transfer. The on-site off-diagonal order parameter $\langle S_{\pm}^2 \rangle$ is in fact the local superconducting order parameter (modulus and phase), it is nonzero only for the on-site $|\pm 1\rangle$ superpositions.

The pseudospin formalism makes it possible to describe the one- and two-particle transport effects in most general form, as well as effects of local and nonlocal correlations in the charge triplet systems\,\cite{Moskvin2013}. The effective Hamiltonian, which does commute with the $z$-component of the total pseudospin $\sum_i S_{iz}$ thus conserving the total charge of the system, can be written as a sum of potential and kinetic energies as follows:
\begin{equation}\label{H}
H = H_{pot} + H_{kin}^{(1)} + H_{kin}^{(2)} ,
\end{equation}
\begin{equation}\label{H-pot}
H_{pot} = \sum_i (\Delta S_{iz}^2 - \mu S_{iz}) + \frac{1}{2}\, V \sum_{<ij>} S_{iz} S_{jz} ,
\end{equation}
\begin{equation}\label{H-kin}
H_{kin}^{(1)} = -\frac{1}{2} \sum_{<ij>} \Big[t^p P_{i+}P_{j-} + t^n N_{i+} N_{j-} + \frac{t^{pn}}{2} (P_{i+} N_{j-} + N_{i+} P_{j-}) + {h.c.}\Big] ,
\end{equation}
\begin{equation}\label{H-tb}
H_{kin}^{(2)} = - \frac{1}{2}\, t^b  \sum_{<ij>} (S_{i+}^2 S_{j-}^2 + S_{i-}^2 S_{j+}^2) .
\end{equation}

With the exception of some $ST$ terms in  (\ref{H-kin}) that are not invariant under the time reversal symmetry, this Hamiltonian is one of the most general anisotropic S\,=\,1  spin-Hamiltonians.  The first term in (\ref{H-pot}), or ``single-ion anisotropy'', describes the on-site density-density correlation effects. The second term can be related to a pseudo-magnetic field along the $z$-axis, or to a chemical potential for doped particles. The last term describes the inter-site density-density interactions. The Hamiltonian (\ref{H-kin}) describes an one-particle hopping process in the system; the PP-term corresponds to $| $Cu$^{2+} \rangle + | $Cu$^{3+} \rangle \leftrightarrow | $Cu$^{3+} \rangle + | $Cu$^{2+} \rangle$ transitions, the NN transport corresponds to $| $Cu$^{2+} \rangle + | $Cu$^{1+} \rangle \leftrightarrow | $Cu$^{1+} \rangle + | $Cu$^{2+} \rangle$, and the PN-term defines a very different one-particle hopping process $| $Cu$^{2+} \rangle + | $Cu$^{2+} \rangle \leftrightarrow | $Cu$^{3+} \rangle + | $Cu$^{1+} \rangle$, that is the local disproportionation/recombination, or the electron-hole pair creation/annihilation. The Hamiltonian (\ref{H-tb}) describes a two-particle (local composite boson) inter-site hopping $| $Cu$^{1+} \rangle + | $Cu$^{3+} \rangle \leftrightarrow | $Cu$^{3+} \rangle + | $Cu$^{1+} \rangle$.

Depending on the relation between parameters of the Hamiltonian (\ref{H}) and the value of the total charge, the ground state corresponds to either a homogeneous nonconducting phase such as the quantum paramagnet (QPM) with $\langle S_z\rangle$\,=\,$\langle S_z^2\rangle$\,=0, which is implemented for large positive values of the correlation parameter  $\Delta$, or a nonconducting phase of the charge ordering (CO) to be analogue of the (anti)ferromagnetic ordering along the $z$-axis, or variants of the superconducting XY-(SF, superfluid) phases with a non-zero order parameters $\langle S_{\pm}\rangle$ and/or  $\langle S_{\pm}^2\rangle$, which can be accompanied by a ferromagnetic or staggered antiferromagnetic order (supersolid phase) for the $z$-component of the pseudospin.

\section{Schwinger boson method}
\label{sec:2}

For the analysis of the S\,=\,1 pseudospin system we made  use of the Schwinger boson representation in the mean field\,\cite{Auerbach1988}. In this method, the three S\,=\,1 pseudospin projections are assigned to the three Bose operators of the quasiparticle creation/annihilation  over vacuum: $|1 \rangle = b^\dagger_+ |v \rangle$, $|0 \rangle = b^\dagger_0 |v \rangle$, $ |-1 \rangle = b^\dagger_- |v \rangle$, with a constraint:
\begin{equation}\label{constraint}
b^\dagger_+ b_+ + b^\dagger_0 b_0 + b^\dagger_- b_- = 1.
\end{equation}

After replacing the pseudospin operators in (\ref{H}) by boson operators: $ P_+ = -b^\dagger_+ b_0$, $ P_- = -P_+^\dagger = b_0^\dagger b_+$, $N_+ = -b^\dagger_0 b_-$, $ N_- = -N_+^\dagger = b_-^\dagger b_0$, $S_z = b^\dagger_+ b_+ - b^\dagger_- b_-,\ S_z^2 = b^\dagger_+ b_+ + b^\dagger_- b_-$, $S_+^2 = b^\dagger_+ b_-$, $S_-^2 = (S_+^2)^\dagger= b_-^\dagger b_+ $, we obtain the Hamiltonian as follows:
$$
H_{pot} = \Delta \sum_{i} \left( b_{i+}^\dagger b_{i+} + b_{i-}^\dagger b_{i-} \right) - \mu \sum_{i} \left( b_{i+}^\dagger b_{i+} - b_{i-}^\dagger b_{i-} \right) +  
$$
$$
+ \frac{V}{2} \sum_{<ij>} \left( b_{i+}^\dagger b_{i+} - b_{i-}^\dagger b_{i-} \right) \left( b_{j+}^\dagger b_{j+} - b_{j-}^\dagger b_{j-} \right) -
$$
\begin{equation}
- \nu \sum_{i} \left( b_{i+}^\dagger b_{i+} + b_{i-}^\dagger b_{i-} + b_0^2 - 1 \right) , 
\end{equation}
\begin{equation}
H_{kin}^{(1)} = \frac{b_0^2}{2} \sum_{<ij>} \left[ t^{p} b_{i+}^\dagger b_{j+} + t^{n} b_{i-}^\dagger b_{j-} + \frac{t^{pn}}{2} \, ( b_{i+}^\dagger b_{j-}^\dagger + b_{i-}^\dagger b_{j+}^\dagger ) + {h.c.} \right],
\end{equation}
\begin{equation}
H_{kin}^{(2)} = - \frac{t^b}{2} \sum_{<ij>} \left( b_{i+}^\dagger b_{i-} b_{j-}^\dagger b_{j+} + b_{i-}^\dagger b_{i+} b_{j+}^\dagger b_{j-} \right) ,
\end{equation}
where we assume the ground (vacuum) state of the quantum paramagnet, which corresponds to the $b_0$ Bose condensate, i.e. $\langle b_0 \rangle = \langle b_0^\dagger \rangle = b_0$. The parameter $\nu$ implements the constraint (\ref{constraint}) in the mean field. The quadratic terms in the Hamiltonian are linearized in the mean field approximation:
\begin{equation}
b_{i+}^\dagger b_{i-} b_{j-}^\dagger b_{j+} + b_{i-}^\dagger b_{i+} b_{j+}^\dagger b_{j-} =  q \, ( b_{i+}^\dagger b_{i-} + b_{j-}^\dagger b_{j+} + {h.c.}) - 2 q^2 , 
\end{equation}
$$
(b^\dagger_{i+} b_{i+} - b^\dagger_{i-} b_{i-}) (b^\dagger_{j+} b_{j+} - b^\dagger_{j-} b_{j-}) =
$$
$$
= \frac{1}{2} (1 - b^2_0 + m) (b^\dagger_{i+} b_{i+} + b^\dagger_{j+} b_{j+}) + \frac{1}{2} (1 - b^2_0 - m) (b^\dagger_{i-} b_{i-} + b^\dagger_{j-} b_{j-}) -
$$
\begin{equation}
- p \, (b_{i+} b_{j-} + b_{i-} b_{j+} + {h.c.}) + 2 p^2 - \frac{1}{2} (1 - b^2_0)^2 - \frac{1}{2} m^2 ,
\end{equation}
where $q = \langle b_{i-}^\dagger b_{i+} \rangle = \langle b_{i+}^\dagger b_{i-} \rangle$, $p = \langle b_{i+}^\dagger b_{j-}^\dagger \rangle = \langle b_{i+} b_{j-} \rangle$, $m = \langle b_{i+}^\dagger b_{i+} \rangle - \langle b_{i-}^\dagger b_{i-} \rangle$.

After a Fourier-Bogoliubov transformation, we get the diagonalized Hamiltonian as follows
\begin{equation}
H = \sum_{\textbf{k}\alpha} \Omega_{\textbf{k}\alpha} B_{\textbf{k}\alpha}^\dagger B_{\textbf{k}\alpha} + \frac{1}{2} \sum_{\textbf{k}\alpha} ( \Omega_{\textbf{k}\alpha} - \Lambda_{\textbf{k}} ) + N C \, ,
\end{equation}
where the following notation is used:
$$
\Omega_{\textbf{k}\alpha} = \sqrt{\omega_{\textbf{k}}^2 + \lambda_{\textbf{k}}^2 + \tau^2 + 2 \kappa_{\textbf{k}\alpha}}  \ ,  \ \alpha=\pm ,
$$
$$
\omega_{\textbf{k}} = \sqrt{\Lambda_{\textbf{k}}^2 - D_{\textbf{k}}^2} \ , \ \kappa_{\textbf{k}\pm} = \pm \sqrt{\omega_{\textbf{k}}^2 \lambda_{\textbf{k}}^2 + \Lambda_{\textbf{k}}^2 \tau^2} ,
$$
$$
\Lambda_{\textbf{k}} = -\nu + \Delta + \frac{1}{2}ZV\,(1 - b^2_0) - Z\, t^m \, b^2_0 \gamma_{\textbf{k}} \ , 
$$
$$
\lambda_{\textbf{k}} = \mu - \frac{1}{2}ZVm + Z\, t^l \, b^2_0 \gamma_{\textbf{k}} \ , 
$$
$$
D_{\textbf{k}} = - \Big( \frac{t^{pn} b^2_0}{2} + V p \Big) Z \gamma_{\textbf{k}} \ , \ \tau = - Z t^b q \ , 
$$
$$
C = \nu (1 - b^2_0) - \frac{1}{4} ZV (1 - b^2_0)^2 - \frac{1}{4} Z V m^2 + Z V p^2 +Z t^b q^2 \ , 
$$
\begin{equation}
\gamma_{\textbf{k}} = \frac{1}{Z} \sum_{<\textbf{r}>} e^{i \textbf{kr}} \ ,\ t^m=\frac{t^p + t^n}{2} \ ,\ t^l=\frac{t^p - t^n}{2} \ . 
\end{equation}

The novel mean field parameters $b_0$, $\nu$, $q$, $p$, $m$ are found from the condition of the free energy minimum $F = N e_0 - \frac{1}{\beta} \sum_{\textbf{k}} \ln [1 + n(\Omega_{\textbf{k}-})] - \frac{1}{\beta} \sum_{\textbf{k}} \ln [1 + n(\Omega_{\textbf{k}+})]$, where $e_0 = \frac{1}{2N} \sum_{\textbf{k}\alpha} (\Omega_{\textbf{k}\alpha} - \Lambda_{\textbf{k}}) + C$ , $n(\Omega_{\textbf{k}\alpha}) = 1/(\exp{\beta \Omega_{\textbf{k}\alpha}} - 1)$.
After minimization, we obtain a system of the self-consistent equations as follows:
$$
2 - b_0^2 = \frac{1}{N} \sum_{\textbf{k}\alpha} \Lambda_{\textbf{k}}
\Big( 1 + \frac{\lambda_{\textbf{k}}^2 +\tau^2}{\kappa_{\textbf{k}\alpha}} \Big) \frac{ n(\Omega_{\textbf{k}\alpha}) + \frac{1}{2}}{\Omega_{\textbf{k}\alpha}} ,
$$
$$
\nu =  \frac{Z}{N} \sum_{\textbf{k}\alpha} \gamma_{\textbf{k}}
\bigg[ \frac{t^{pn}}{2}D_{\textbf{k}} - t^m \Lambda_{\textbf{k}} + t^l \lambda_{\textbf{k}} +
$$
$$
+\frac{ (\frac{t^{pn}}{2}D_{\textbf{k}} - t^m \Lambda_{\textbf{k}} ) \lambda_{\textbf{k}}^2 + t^l \omega_{\textbf{k}}^2 \lambda_{\textbf{k}} - t^m \tau^2 \Lambda_{\textbf{k}} }{\kappa_{\textbf{k}\alpha}} \bigg] \frac{ n(\Omega_{\textbf{k}\alpha}) + \frac{1}{2}}{\Omega_{\textbf{k}\alpha}} ,
$$
$$
q = \frac{1}{2N} \sum_{\textbf{k}\alpha} \tau  
\Big( 1 + \frac{\Lambda_{\textbf{k}}^2}{\kappa_{\textbf{k}\alpha}} \Big) \frac{ n(\Omega_{\textbf{k}\alpha}) + \frac{1}{2}}{\Omega_{\textbf{k}\alpha}} , 
$$
$$
p = - \frac{1}{2N} \sum_{\textbf{k}\alpha} D_{\textbf{k}} \gamma_{\textbf{k}} 
\Big( 1 + \frac{\lambda_{\textbf{k}}^2}{\kappa_{\textbf{k}\alpha}} \Big) \frac{ n(\Omega_{\textbf{k}\alpha}) + \frac{1}{2}}{\Omega_{\textbf{k}\alpha}} ,
$$
\begin{equation}\label{sys}
m = - \frac{1}{N} \sum_{\textbf{k}\alpha} \lambda_{\textbf{k}}
\Big( 1 + \frac{\omega_{\textbf{k}}^2}{\kappa_{\textbf{k}\alpha}} \Big) \frac{ n(\Omega_{\textbf{k}\alpha}) + \frac{1}{2}}{\Omega_{\textbf{k}\alpha}} .
\end{equation}

\section{Results}
\label{sec:3}
We made calculations of the pseudospin excitations, or pseudomagnons, on a 512$\times$512 square lattice with free boundary conditions. A numerical solving of equations (\ref{sys}) gives the conditions when the energy gap in the pseudomagnon spectrum goes to zero, indicating a phase transition in the system.
\begin{figure}[H]
  \includegraphics[width=0.45\textwidth]{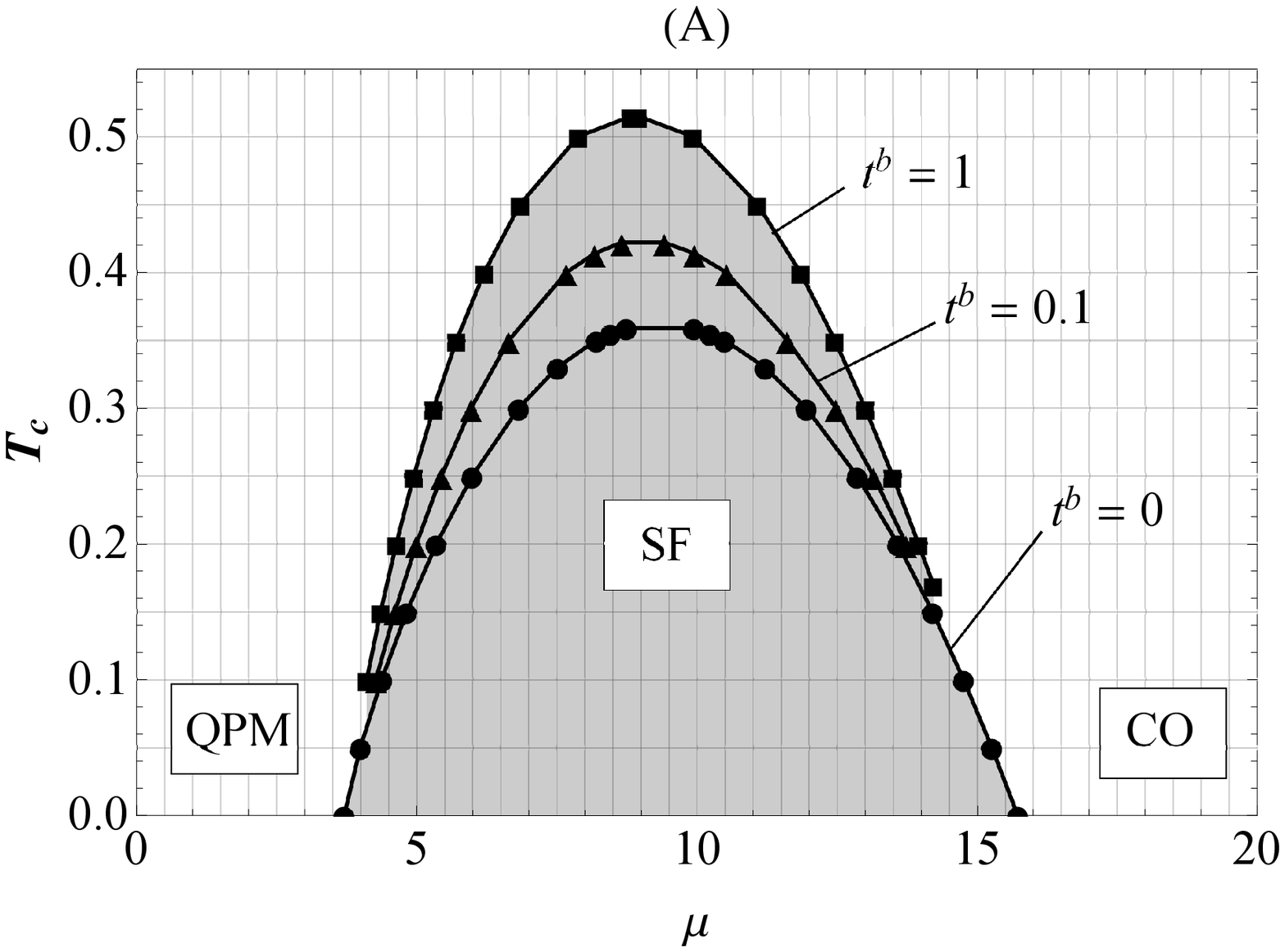}
  \includegraphics[width=0.45\textwidth]{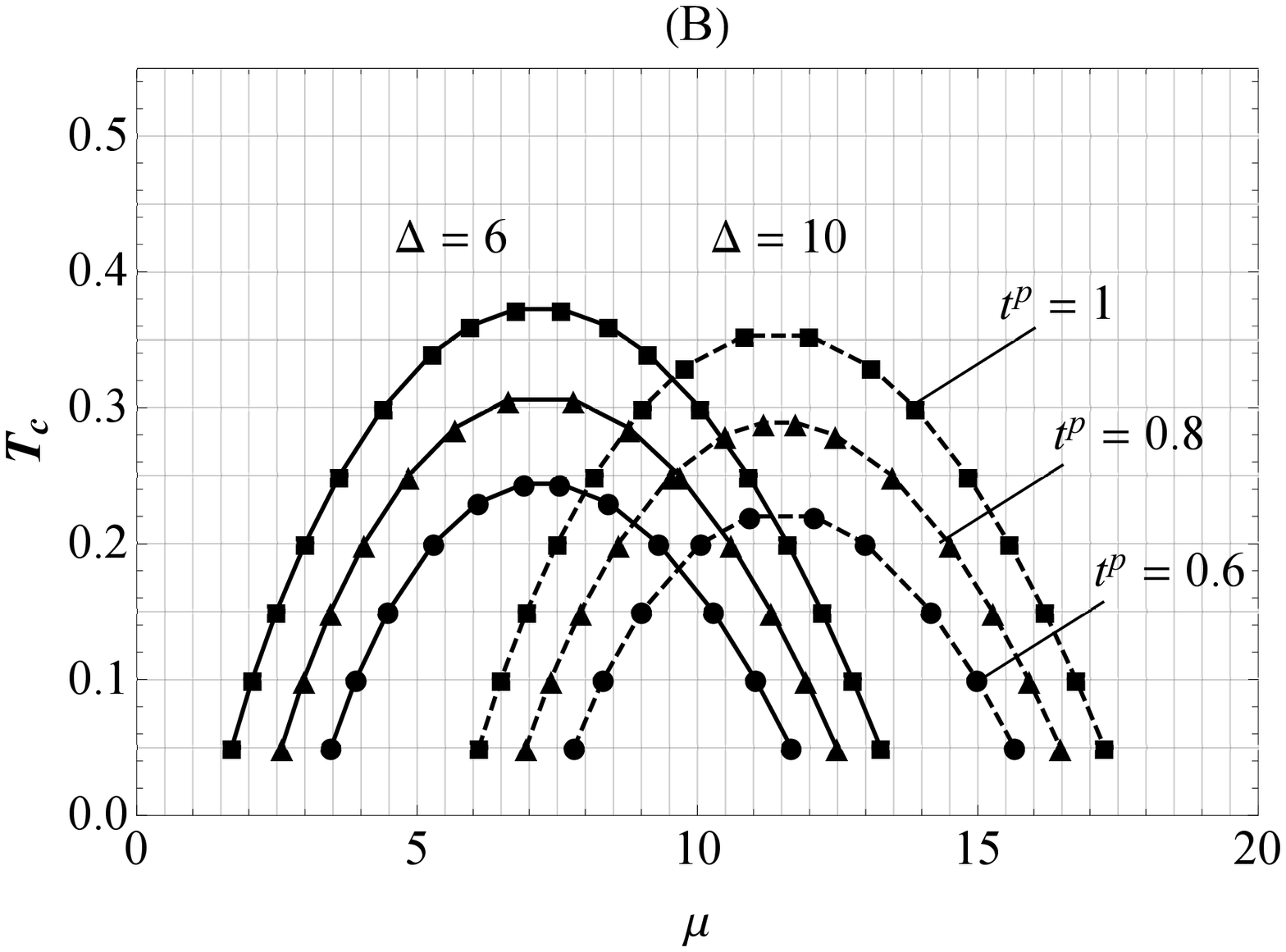}
\caption{$T - \mu$ phase diagram for $V=t^{n}=t^{pn}=1$. QPM~--- quantum paramagnet, SF~--- superfluid, CO~--- charge oder. (A): the effect of 2-particle transport $t^b$ for $\Delta=8$, $t^p=1$. (B): the effects of 1-particle transport $t^p$ and the local energy of $M^{0,\pm}$ centers $\Delta$ for $t^b = 0$.}
\label{fig:1}
\end{figure}
\begin{figure}[H]
  \includegraphics[width=0.44\textwidth]{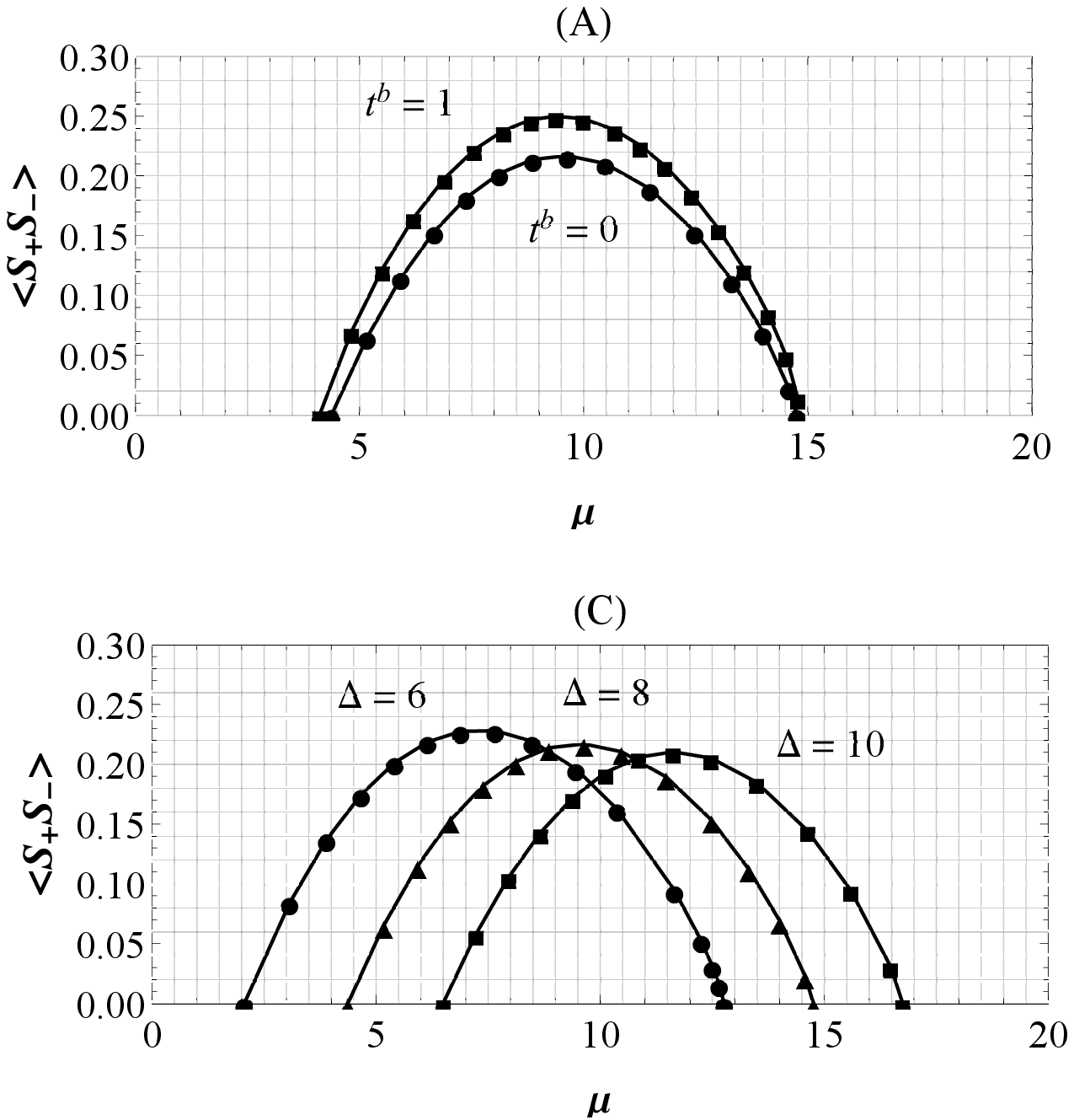}
  \includegraphics[width=0.45\textwidth]{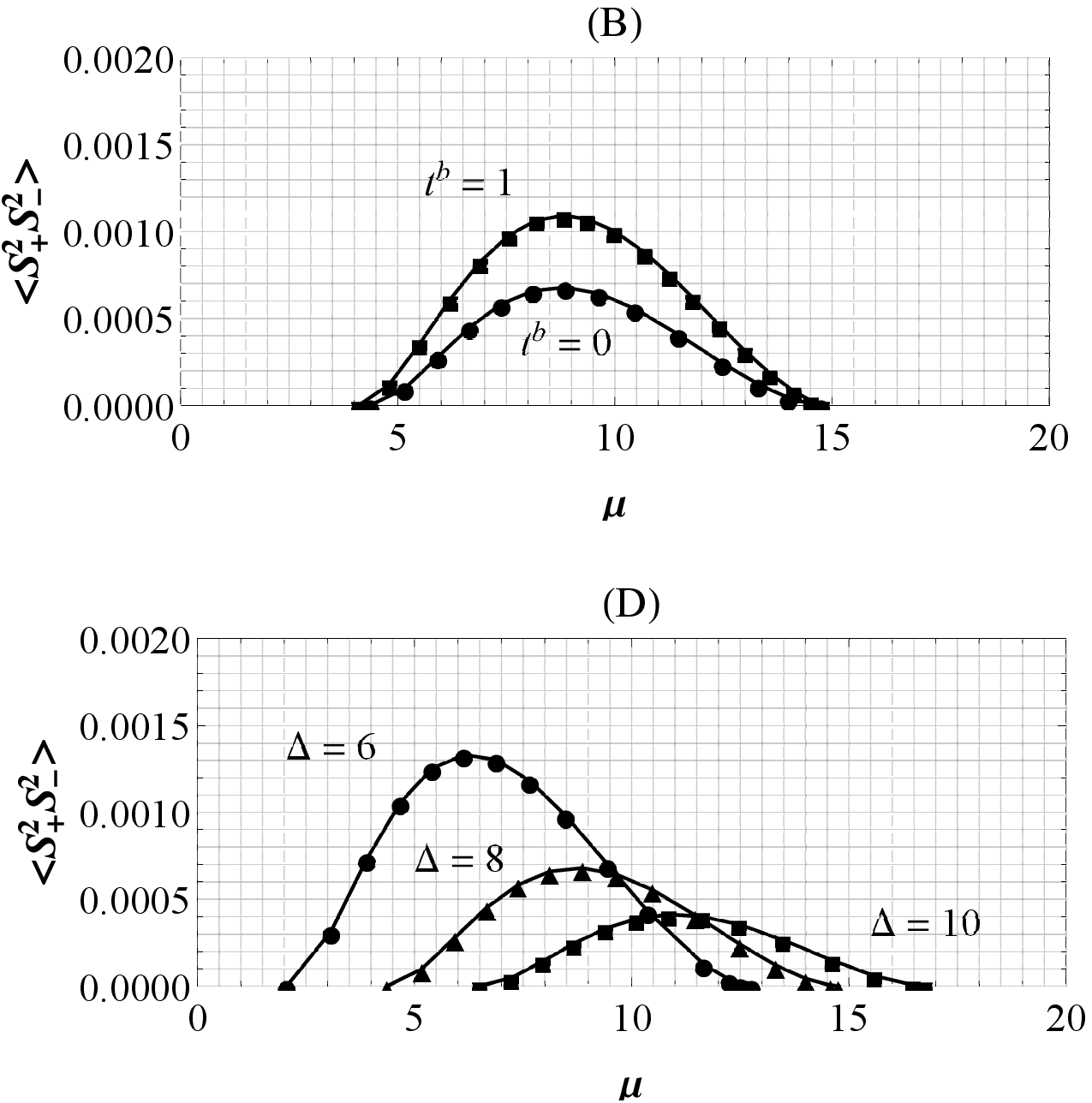}
\caption{The correlators $\langle S_{+} S_{-} \rangle$ (A) and $\langle S_{+}^2 S_{-}^2 \rangle$ (B) for fig. 1a and the correlators $\langle S_{+} S_{-} \rangle$ (C) and $\langle S_{+}^2 S_{-}^2 \rangle$ (D) for fig. 1b. at $T=0.01$ and $t^p=1$.}
\label{fig:2}
\end{figure}

The excitations split in the pseudomagnetic field (chemical potential $\mu$) and the component $\Omega_{\textbf{k}-}$ decreases with the increasing parameter $\mu$. At a critical value of potential $\mu_{c1}$, the energy gap disappears. When the pseudomagnetic field further increases, we assume the energy gap keeps zero and part of the excitations condense either at the point $\textbf{k} = (\pi , \pi)$ in case $t^p > 0$, or at the point $\textbf{k} = (0 , 0)$ in case $t^p < 0$. Consequently, the magnetization $m$ parallel to the $z$-axis appears and at the same time a magnetization in the plane occurs. At a second critical value of potential $\mu_{c1}$ the magnetization $m$ saturates and the plane magnetization disappears. For a given pseudomagnetic field $\mu_{c1} < \mu < \mu_{c2}$, a critical temperature $T(\mu)$ exists, below which, the energy gap keeps zero and part of the excitations are condensed.

Thus, Fig.\,\ref{fig:1} shows the phase transitions between the quantum paramagnet (parent cuprate Cu$^{2+}$), the spin-flop state (superfluidity) and the ferromagnet (charge-ordered phase Cu$^{3+}$). As can be seen in Fig.\,\ref{fig:2}, the SF phase includes both the single- and two-particle superfluidity (i.e. superconductivity). The parameter $t^b$ in this approach though has the effect to the two-particle transport, but it is rather weak, its main effect is in the increase of the critical temperatures $T_c$. The local energy of the $M_{0,\pm}$ centers $\Delta$ (single-ion anisotropy) gives the expected results: its rise leads to suppression of the order parameter $\langle S_{\pm}^2 \rangle$ and shift of the SF phase in the region of large pseudomagnetic fields.
\begin{figure}[H]
  \includegraphics[width=0.32\textwidth]{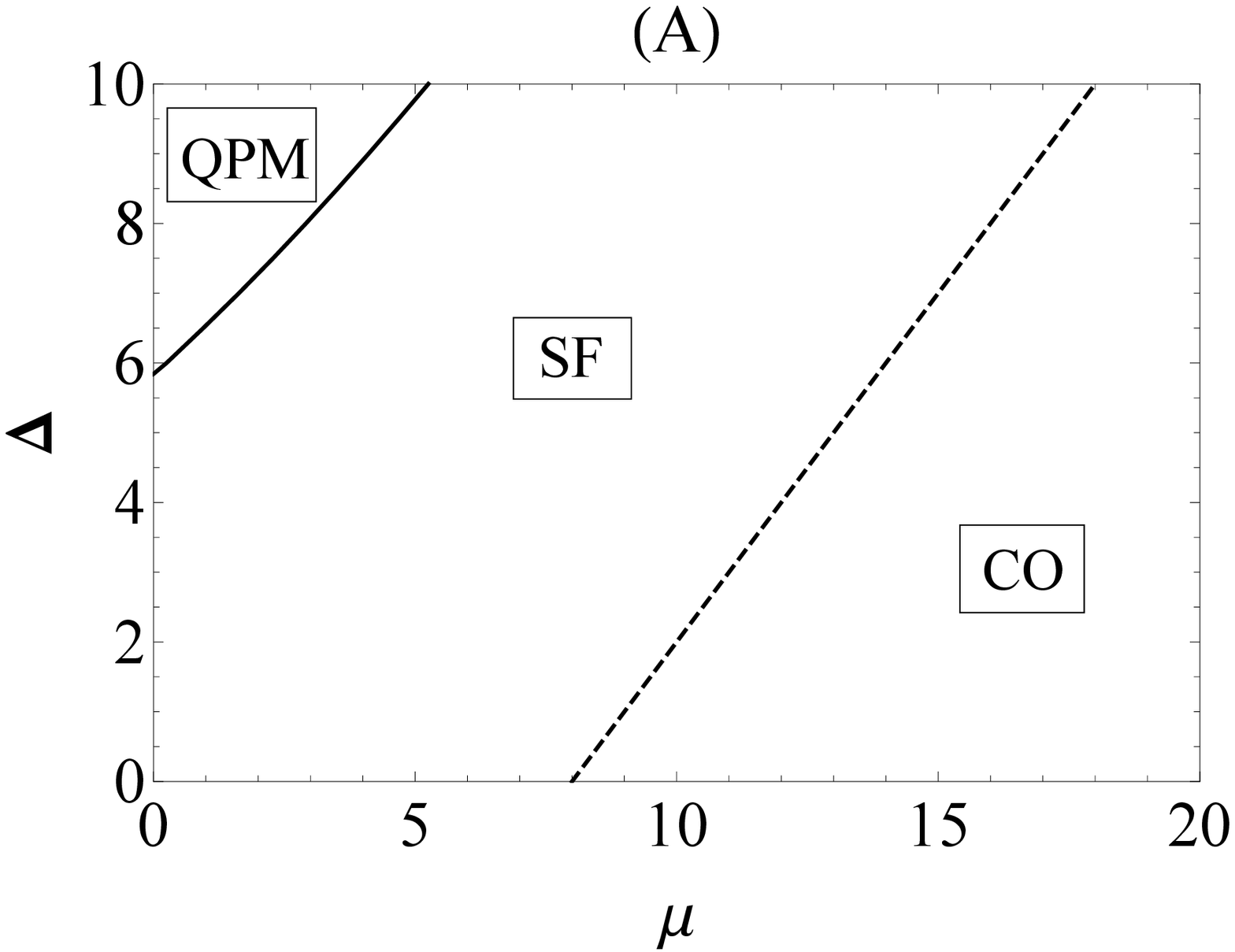}
  \includegraphics[width=0.32\textwidth]{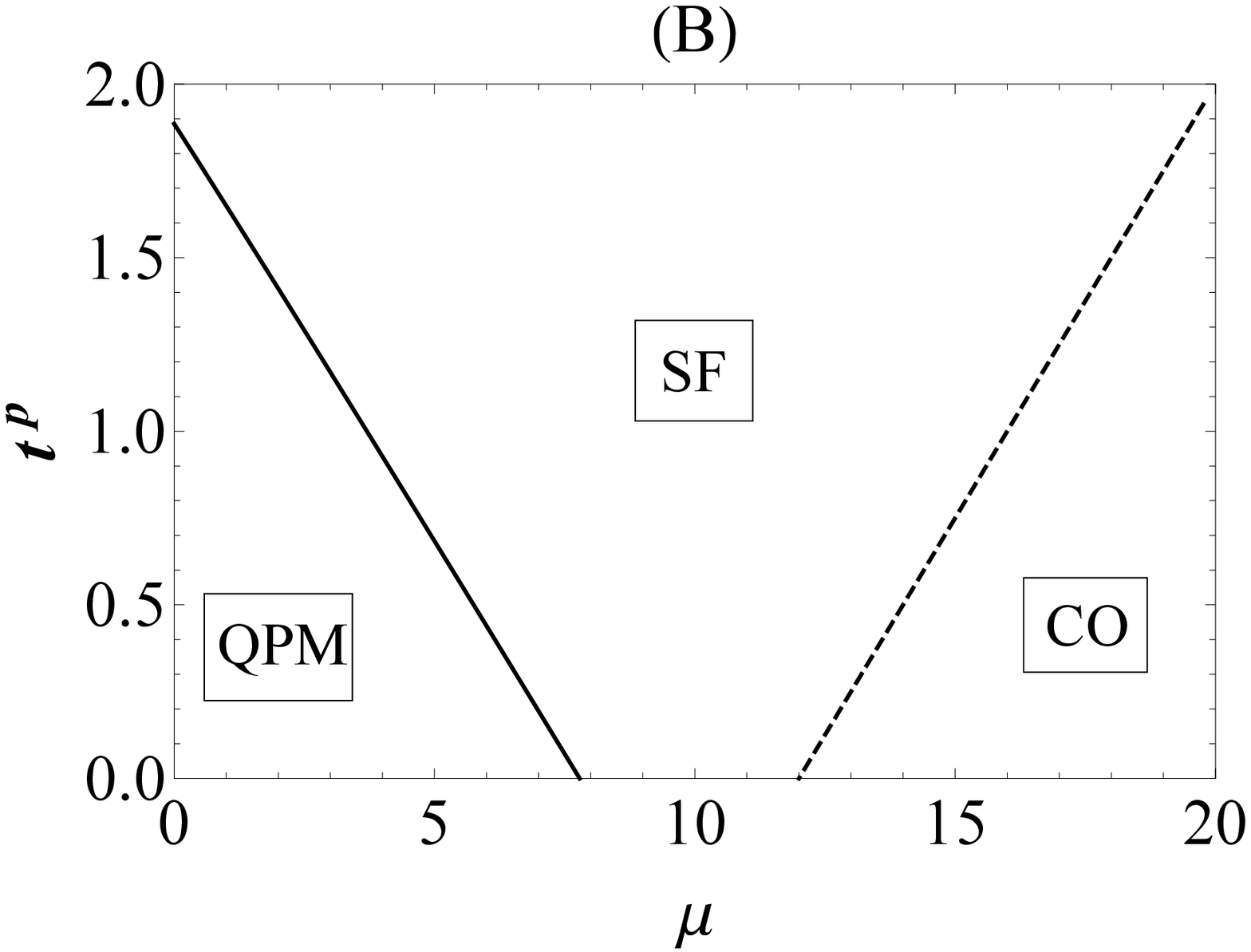}
  \includegraphics[width=0.32\textwidth]{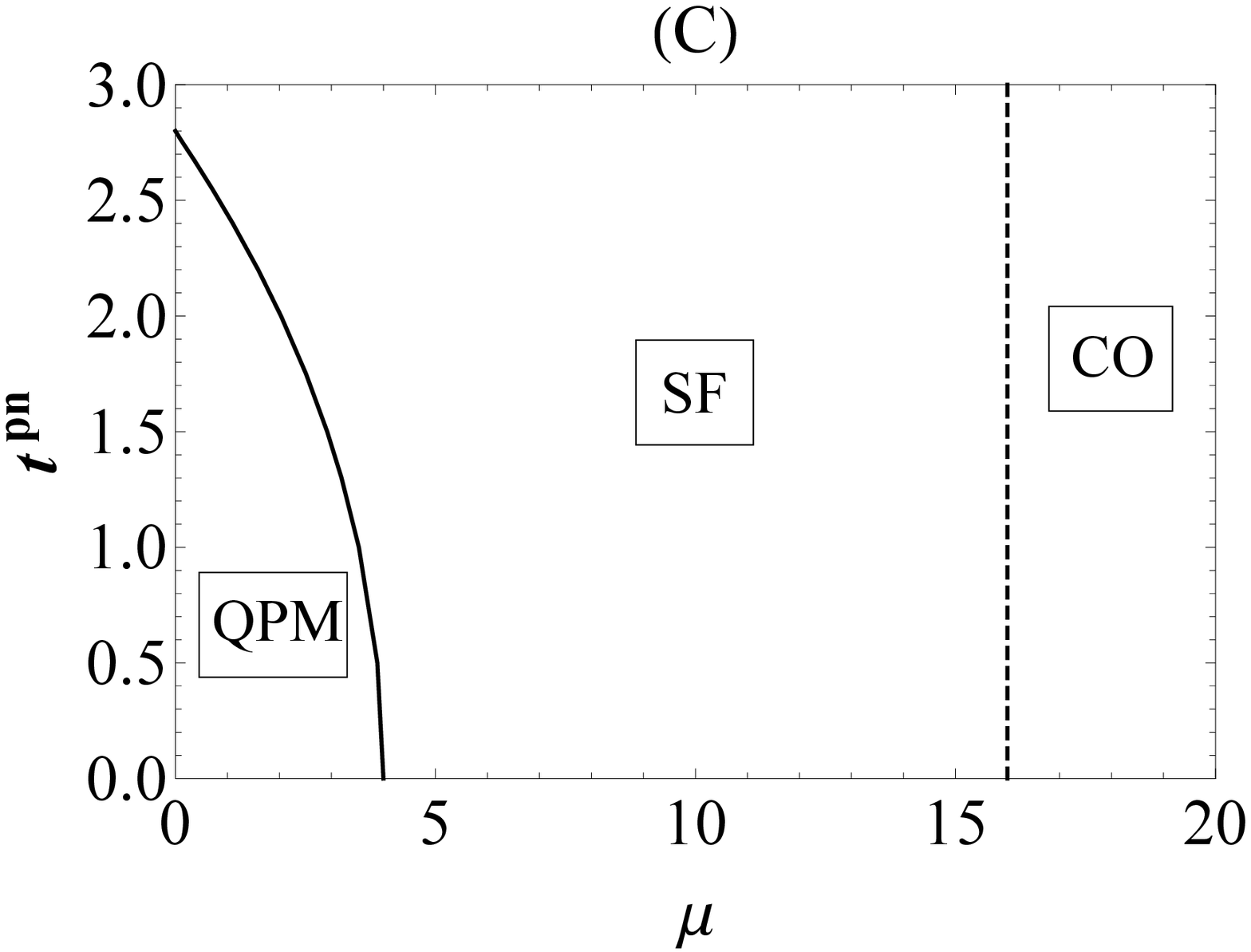}
\caption{The ground state phase diagrams for $V=t^{n}=1$ in coordinates $\Delta - \mu$ (A), $t^p - \mu$ (B) and $t^{pn} - \mu$ (C).}
\label{fig:3}
\end{figure}

The main contribution to the formation of the SF phase in the ground state (Fig.\,\ref{fig:3}b) and to the value of the critical temperatures (Fig.\,\ref{fig:1}b) is made by the PP-type single-particle transport. The NN transport has the least effect, but this depends on the geometry of the problem: if the field $\mu$ is directed against the $z$-axis (electron doping), $t^n$ will change roles with $t^p$.

Interestingly, the PN transfer (local disproportionation) can form the SF phase, that's why it does not disappear in the Fig.\,\ref{fig:3}b at $t^p = 0$, but unlike the PP transport at $t^{pn} = 0$ we don't observe the superconductivity ($| $Cu$^{1+} \rangle + | $Cu$^{3+} \rangle \leftrightarrow | $Cu$^{3+} \rangle + | $Cu$^{1+} \rangle$), which is due to the fact that we are essentially removing the mechanism of the Cu$^{1+}$ centers creation. In addition, when the saturation $\mu_{c2}$ is reached, the local disproportionation effects also can not appear, which agrees with the data in Fig.\,\ref{fig:3}c.

\section{Conclusions}
\label{sec:4}
Within the framework of the S\,=\,1 pseudospin formalism, we investigated the model high-$T_c$ cuprate with the three effective valence centers Cu$^{1+,\,2+,\,3+}$ or the system of ``semi-hard-core'' bosons with the constraint on the lattice site occupancy n = 0, 1, 2. Starting with a parent cuprate as an analogue of the quantum paramagnetic ground state and using the Schwinger boson technique we found the pseudomagnon dispersion relations and conditions for the pseudomagnon condensation with phase transition to the superconducting state. We found that the SF phase includes both a one- and two-particle superfluidity. Single-particle component of the SF phase is determined mainly by the PP-type transfer for the hole doping or by the NN-type transfer for the electron doping. Two-particle component of the SF phase is determined mainly by the PN-type transport and in lesser degree by the two-particle  $t^b$ term.

\begin{acknowledgements}
The research was supported by the Government of the Russian Federation, Program 02.A03.21.0006, by the Ministry of Education and Science of the Russian Federation, Projects nos. 2277 and 5719, and by the Competitiveness Enhancement Program - CEP 3.1.1.2-18.
\end{acknowledgements}

\end{document}